\documentstyle[aps,epsf]{revtex}

\newcommand{\be}{\begin{equation}}
\newcommand{\ee}{\end{equation}}
\newcommand{\bea}{\begin{eqnarray}}
\newcommand{\eea}{\end{eqnarray}}

\begin{document}
\draft\onecolumn
\title{Optimal universal quantum cloning and state estimation}
\author{Dagmar~Bru\ss $^1$, Artur~Ekert$^2$,
Chiara~Macchiavello$^{3,1}$}
\address{
$^1$ISI, Villa Gualino, Viale Settimio Severo 65, 10133 Torino, Italy\\
$^2$Clarendon Laboratory, University of Oxford, Parks Road, Oxford OX1 3PU,\\
UK\\
$^3$Dipartimento di Fisica ``A. Volta'' and I.N.F.M., \\
Via Bassi 6, 27100 Pavia, Italy}
\date{Received \today}
\maketitle
\begin{abstract}
We derive a tight upper bound for the fidelity of a universal $N\to M$ qubit
cloner, valid for any $M\geq N$,
where the output of the cloner is required to be supported on the 
symmetric subspace.
 Our proof is based on the concatenation of
two cloners and the connection between quantum cloning and quantum state
estimation. 
We generalise the operation of a quantum cloner to mixed and/or entangled 
input qubits described by a density matrix supported on the
symmetric subspace of the constituent qubits. We also extend the validity of
optimal state estimation methods to inputs of this kind.
\end{abstract}
Perfect quantum cloning is impossible~\cite{Wootters}. This notwithstanding,
we may ask how well we can clone quantum states. Bu\v{z}ek and Hillery, who
were the first to address this problem, provided an example of a quantum
device which can clone an unknown pure state of a single input qubit (a
two-state system) into two output qubits,~albeit with a certain fidelity
smaller than one\cite{vlado}. Their construction was subsequently shown to
be optimal~by Bru\ss\ {\it et al}. \cite{oxibm}. In this letter we derive
the optimal fidelity of a universal and symmetric quantum cloning machine
(QCM) which acts on $N$ original qubits and generates $M$ clones.

A universal $N\rightarrow M$ quantum cloner is a quantum machine which
performs a prescribed unitary transformation on an extended input which
contains $N$ original qubits, $M-N$ ``blank'' qubits and $K$ auxiliary
qubits, and which outputs $M$ clones together with the $K$ auxiliary qubits.
The original qubits are all in the same (unknown and pure) quantum state
described by the density operator $\varrho ^{in}=\frac{1}{2}(%
\mbox{$1%
\hspace{-1.0mm}  {\bf l}$}+{\vec{s}}^{\,in}\cdot \vec{\sigma})$, where ${%
\vec{s}}^{\,in}$ is the original Bloch vector. Both ``blanks'' and the
auxiliary qubits are initially in some prescribed quantum state. The output
qubits are in an entangled state and in the present work
we require that the density operator
describing the state of the $M$ clones is supported on the symmetric
subspace. 
This guarantees that all the output qubits are indistinguishable and in 
the same state described by the reduced density operator
 $\varrho ^{out}$. We comment on relaxing this assumption at the end of the 
paper.
\par
 It has been shown that {\em universal} $1\to 2$ cloners
can only shrink the original Bloch vector, without changing its orientation
in the Bloch sphere \cite{oxibm}. 
The same argument as given in \cite{oxibm} (namely the impossibility
to find a transformation that rotates
 {\em any} Bloch vector of the 1-particle reduced density matrix
by the same angle) applies also generally for an $N\to M$ qubit cloner.
Therefore, the operation of a universal
QCM can be characterised by the shrinking factor $\eta (N,M)$, 
which is also known in the literature as 
the Black Cow factor \cite{Werner}, 
and the
reduced output density operator is of the form $\varrho ^{out}=\frac{1}{2}(%
\mbox{$1 \hspace{-1.0mm}  {\bf l}$}+\eta (N,M){\vec{s}}^{\,in}\cdot \vec{\sigma})$. Universal $N\rightarrow M$ quantum cloning machines may be
constructed in many different ways, the best constructions are those which
maximize $\eta (N,M)$ (i.e. which maximize the fidelity of the cloning
machine) and we refer to them as the optimal cloners.  

Gisin and Massar have constructed a class of universal $N\rightarrow M$ QCMs
and showed, using numerical methods, that for $N\leq 7$ their cloners are
optimal~\cite{gima}. Our derivation of the upper bound for $\eta (N,M)$ is
quite general and does not refer to any specific realisation of the
universal cloning machines. In particular it shows that the Gisin-Massar
cloners saturate this bound for any $N$ and $M\geq N$. Our approach avoids
an elaborate optimisation procedure, extends the class of allowed inputs to
mixed and/or entangled states of the original qubits which belong to the
symmetric subspace and sheds some light on the connection between optimal
quantum cloning and optimal quantum 
state estimation. The proof is based on the
concatenation of two quantum cloners and on associating the upper bound on
the fidelity of an $M\rightarrow \infty $ cloner with the fidelity of the
optimal state estimation of $M$ qubits, given in \cite{mapo}.

We concatenate two cloning machines in the following way. The first cloner
is an $N\to M$ universal machine characterised by the shrinking factor $%
\eta(N,M)$. The $M$ clones from the output of the first cloner are then
taken as originals for the input into the second cloning machine which
creates infinitely many clones with the shrinking factor $\eta(M,\infty)$.
We now write down two statements which will be proved after unfolding the
main result:

\begin{itemize}
\item[a)]  The shrinking factors for concatenated cloners multiply.

\item[b)]  The equality
\begin{equation}
\eta _{QCM}^{opt}(M,\infty )={\bar{\eta}}_{meas}^{opt}(M)
\end{equation}
holds. Here ${\bar{\eta}}_{meas}^{opt}(M)$ corresponds to the optimal state
estimation derived in \cite{mapo}, and its meaning will be explained below.
\end{itemize}

Due to statement a), the shrinking factors of universal cloning machines in
sequence multiply. Moreover, the sequence of the two machines cannot perform
better than the optimal $N\rightarrow \infty $ universal cloner, otherwise
the $N\rightarrow \infty $ universal cloner would not be optimal. Thus we
arrive at the following inequality:
\begin{equation}
\eta _{QCM}(N,M)\cdot \eta _{QCM}(M,\infty )\leq \eta _{QCM}^{opt}(N,\infty
)\ \ .  \label{conc-a}
\end{equation}
This means that the lowest upper bound for the general $N\rightarrow M$
cloner is given by
\begin{equation}
\eta _{QCM}(N,M)\leq\frac{\eta _{QCM}^{opt}(N,\infty )}{\eta
_{QCM}^{opt}(M,\infty )}\ \ .  \label{conc} 
\end{equation}
We have thus reduced the optimality problem of the $N\rightarrow M$ cloner
to the task of finding the optimal $N\rightarrow \infty $ cloner.

Now we can use statement b) and the explicit form of $\bar\eta%
_{meas}^{opt}(M)$ (see~\cite{mapo}), namely

\begin{equation}
\bar\eta^{opt}_{meas} (M)=\frac{M}{M+2} \ \   \label{etameas}
\end{equation}
to conclude the central result that for any $M\geq N$
\begin{equation}
\eta_{QCM}^{opt}(N,M)=\frac{N}{M}\frac{M+2}{N+2}\;.\ \   \label{etanm}
\end{equation}
For pure input states this corresponds to the optimal fidelity
\begin{equation}
F_{QCM}^{opt}(N,M)=\frac{NM+N+M}{M(N+2)} \ \ ,
\end{equation}
which is achieved by the cloning transformations proposed in \cite{gima}.
(For $\varrho^{in}= | \, \psi \rangle \langle \psi \, |$ the fidelity is
defined as $F=\langle\psi|\varrho^{out}|\psi\rangle$.)

Let us note in passing that as the consequence of the factorisation property
(\ref{conc}) we can produce $M$ clones from $N$ originals either by applying
directly the optimal $N\rightarrow M$ cloner or by taking any number of
intermediate steps in order to realise the cloning process, using the
optimal transformation at each step; both ways lead to the same overall
shrinking factor.

Let us now justify statements a) and b).

In order to prove a) we describe an $N\rightarrow M$ cloner in terms of a completely positive map $C$\noindent $_{NM\text{ }}$ which maps input
density operators of $N$ identical pure
originals into output density operators of $M$
clones, such that for any state $|\psi \rangle \langle \psi |$ of a single
input qubit (original) we have

\begin{equation}
{\rm Tr}_{M-1}[C_{NM}(|\psi \rangle \langle \psi |^{\otimes N})]=\eta
(N,M)|\psi \rangle \langle \psi |+(1-\eta (N,M))\frac{1}{2}{\mbox{$1
\hspace{-1.0mm}  {\bf l}$}} \ ,
\label{map}
\end{equation}
where the trace is performed on any $M-1$ qubits
(for an overview of completely positive operators see \cite{kraus}).

Let $\varrho _{N}$ be a density operator of $N$ qubits which is supported on
the symmetric subspace of the $2^{N}$ dimensional Hilbert space. We can
always write  $\varrho _{N}$  as\noindent\ a linear combination of direct
products of identical pure states, $\varrho _{N}=\sum_{i}\alpha _{i}|\psi
_{i}\rangle \langle \psi _{i}|^{\otimes N}$ , where $\sum_{i}\alpha _{i}=1$;
N.B. we do not require that all values  $\alpha _{i}$ are positive 
\cite{Werner,stv}.
The linearity of the completely positive map  and its universality, i.e.
the fact that $\eta (N,M)$ does not depend on $|\psi \rangle $, allow us to extend Eq.(%
\ref{map}) to the more general form

\begin{equation}
{\rm Tr}_{M-1}[C_{NM}(\varrho _{N})]=\eta (N,M)\varrho +(1-\eta (N,M))\frac{1%
}{2}{\mbox{$1 \hspace{-1.0mm}  {\bf l}$}}  \label{cp2}
\end{equation}

where $\varrho ={\rm Tr}_{N-1}[\varrho _{N}]$. Now, suppose we concatenate
an $N\rightarrow M$ and an $M\rightarrow L$ cloner and view it as an $%
N\rightarrow L$ cloner. It evolves the initial $N$ qubit state $\varrho _{N}$
\noindent \noindent first into the $M$ qubit state $\varrho _{M}^{^{\prime }}
$ and then into the $L$ qubit state  $\varrho _{L}^{^{\prime \prime }}$. The
corresponding single qubit reduced density operators are  $\varrho $,  $%
\varrho ^{^{\prime }}={\rm Tr}_{M-1}[\varrho _{M}^{^{\prime }}]$ and $\varrho
^{^{\prime \prime }}={\rm Tr}_{L-1}[\varrho _{L}^{^{\prime \prime }}]$ .
Following Eq.(\ref{cp2}) we can write

\begin{equation}
\varrho ^{^{\prime \prime }}=
\eta (M,L)\varrho ^{^{\prime }}+(1-\eta (M,L))\frac{1}{2}
{\mbox{$1 \hspace{-1.0mm}  {\bf l}$}=}
\eta (N,M)\cdot\eta (M,L)\varrho
+(1-\eta (N,M)\cdot\eta (M,L))\frac{1}{2}{\mbox{$1 \hspace{-1.0mm}  {\bf l}$}}
\end{equation}

i.e. indeed $\eta (N,L)=\eta (N,M)\cdot\eta (M,L)$.

\bigskip

We will now prove statement b). Equation~(\ref{etanm}) was obtained assuming
the following result (due to \cite{mapo}): given $M$ qubits all in an
unknown quantum state $|\,\psi \rangle $ there exists a universal POVM
measurement $\left\{ P_{\mu }\right\} $ \cite{rado} 
which leads to the best possible
estimation of $|\,\psi \rangle $ with fidelity $\bar{F}(M)=\frac{M+1}{M+2}$,
or, equivalently, with
$
\bar{\eta}(M)=\frac{M}{M+2}\ \ .
$
The outcome of each instance of the measurement provides, with probability $%
p_{\mu }(\psi )={\rm Tr}(P_{\mu }|\psi \rangle \langle \psi |^{\otimes M})$,
the ``candidate'' $|\,\psi _{\mu }\rangle $ for $|\,\psi \rangle .$ The
fidelity $\bar{F}_{meas}(M)$ 
is then calculated from the outcomes of \noindent the
measurement as

\begin{equation}
\bar{F}_{meas}(M)=\sum_{\mu }p_{\mu }(\psi )|\langle \psi |\psi _{\mu }\rangle
|^{2}=\langle \psi |\bar{\varrho}|\,\psi \rangle ,
\end{equation}
where $\bar{\varrho}=\sum_{\mu }p_{\mu }(\psi )|\psi _{\mu }\rangle \langle
\psi _{\mu }|$ . In the optimal, universal state estimating procedure the
fidelity must not depend on $\psi $, thus $\bar{\varrho}$ can also be written as

\begin{equation}
\bar{\varrho}=\bar{\eta}_{meas}^{opt}(M)|\psi \rangle \langle \psi |+(1-\bar{%
\eta}_{meas}^{opt}(M))\frac{1}{2}{\mbox{$1 \hspace{-1.0mm}  {\bf l}$}.}
\label{barrho}
\end{equation}

The optimal measurement of this type can be viewed as an $M\rightarrow
\infty $ cloner because after reading each outcome we can prepare any number
of ``candidates'', in particular infinitely many of them, with the average
reconstruction fidelity $\bar{F}_{meas}^{opt}(M)$ with respect to the 
originals. Clearly
this procedure cannot provide a larger shrinking factor than the optimal $%
M\rightarrow L$ cloner and we find
\begin{equation}
\bar{\eta}_{meas}^{opt}(M)\leq \eta _{QCM}^{opt}(M,L)\ \
\label{leq}
\end{equation}
for any $L\geq M$, in particular for $L\rightarrow \infty $.

Let us now show that for $L\rightarrow \infty $ the formula (\ref{leq})
becomes the equality. To see this let us  concatenate an $M\rightarrow L$
cloner with a subsequent optimal state estimating measurement. The input to
the cloner is of the form $|\psi \rangle \langle \psi |^{\otimes M}$ and the
output is described by the density operator $\varrho _{L}$ which is of the
form $\sum_{i}\alpha _{i}|\psi _{i}\rangle \langle \psi _{i}|^{\otimes L}$ ,
where $\sum_{i}\alpha _{i}=1$. 
 The reduced density operator of each output qubit is $\varrho =%
{\rm Tr}_{L-1}\varrho _{L}=\sum_{i}\alpha _{i}|\psi _{i}\rangle \langle \psi
_{i}|=\eta (M,L)|\psi \rangle \langle \psi |+(1-\eta (M,L))\frac{1}{2}{%
\mbox{$1 \hspace{-1.0mm}  {\bf l}$}}$. The cloner $M\rightarrow L$
concatenated with the state estimation on $L$ qubits can be viewed as the
state estimation performed on $M$ qubits.  The total procedure gives the
fidelity of estimating $|\psi \rangle $ which can be written as

\begin{eqnarray}
\bar{F}_{meas}(M) &=&\langle \psi |\sum_{\mu }{\rm Tr}(P_{\mu }\varrho _{L})|\psi _{\mu
}\rangle \langle \psi _{\mu }|\,\psi \rangle =
\sum_{\mu ,i}\langle \psi
|\alpha _{i}{\rm Tr}(P_{\mu }|\psi _{i}\rangle \langle \psi _{i}|^{\otimes
L})|\psi _{\mu }\rangle \langle \psi _{\mu }|\psi \rangle  \\
&=&\sum_{i}\langle \psi |\alpha _{i}[\bar{\eta}_{meas}^{opt}(L)|\psi
_{i}\rangle \langle \psi _{i}|+(1-\bar{\eta}_{meas}^{opt}(L))\frac{1}{2}{%
\mbox{$1 \hspace{-1.0mm}  {\bf l}$}]|}\psi \rangle
\end{eqnarray}
 which for $L\rightarrow \infty $  becomes (due to
$\bar{\eta}_{meas}^{opt}(L)\rightarrow 1$) 
\begin{equation}
\bar{F}_{meas}(M) \to 
\sum_{i}\langle \psi |\alpha _{i}|\psi _{i}\rangle \langle \psi _{i}{|}\psi
\rangle =\langle \psi |\varrho {|}\psi \rangle =\frac{1}{2}(1+\eta
_{QCM}(M,\infty )).
\end{equation}

The concatenation of a cloner with a measurement cannot perform better than
the optimal measurement, 
thus we can write
\begin{equation}
\eta _{QCM}^{opt}(M,\infty )\leq \bar{\eta}_{meas}^{opt}(M)\ \ .
\label{geq}
\end{equation}
Combining equations (\ref{leq}) and (\ref{geq}) finally leads to
\begin{equation}
\eta _{QCM}^{opt}(M,\infty )=\bar{\eta}_{meas}^{opt}(M)\ \ ,
\end{equation}
thus proving statement b).
\par
Before concluding, we want to stress that, as a consequence of what 
was shown
above, 
we can extend the operation of a cloning machine and the application of an
optimal measurement to any input density operator of $N$ qubits which has
support on
the symmetric subspace. The properties of the universal
cloning machine as defined at the beginning of this paper allow us to
conclude that the same machine can operate on {\em any} such symmetric
density operator and shrinks the Bloch vector of the reduced input density
matrix by a fixed amount, independent of the initial length.
Notice also that the optimal machine, for products of pure inputs specified
by the shrinking factor (\ref{etanm}), is still optimal for this extended
class of inputs. Actually, if a better cloning machine existed for mixed
input states, we would use it as the second cloner $M\to\infty$ in Eq. (\ref
{conc-a}), giving a smaller lower bound in Eq. (\ref{conc}). This would lead
to a contradiction because we already know that universal cloners for pure
states saturating the bound (\ref{etanm}) exist \cite{gima}. 
\par
One may want to relax our restriction  and consider 
quantum cloners which produce identical clones (i.e. with the same 
single-qubit density operator $\varrho^{out}$), but for which the state of all 
outputs does not belong to the symmetric subspace.  This case, in principle, 
may provide a higher shrinking factor, however, we could neither prove nor 
disprove this with our approach. We leave this problem as a challenge to other
colleagues.

In a similar way as for the cloner we can extend the validity of an optimal
universal measurement procedure to inputs from the symmetric subspace. In
this case the goal is to find the optimal measurement for the reduced
density operator for each input copy. Since we require the process to be
universal, we know that the reduced density operator reconstructed as the
result of the measurement given in Eq. (\ref{barrho}) is just the shrunk
version of the initial one. We can then describe the quality of the
procedure in terms of the shrinking factor. We conclude that the optimal
measurement derived in Ref. \cite{mapo} is also optimal for any input
symmetric state. Actually, if this were not the case, we could devise a
measurement procedure on $N$ initial pure qubits by first applying an $N\to
M $ cloner and then an optimal measurement on the mixed state of the output $%
M$ clones. If this global measurement were better than the optimal one of
Ref. \cite{mapo} we would then obtain with the above procedure a universal
measurement for pure states which performs better than the one in Ref. \cite
{mapo}, thus finding a contradiction.

Let us mention in passing
 that in our discussion we found it convenient to use
the shrinking factor, because it has an intuitive geometrical meaning both
for pure and mixed states, however, one can rephrase the optimality argument
for universal operations using, for example, the Uhlmann fidelity \cite{uhl}
for the reduced density operators.

In conclusion, we have derived the optimal shrinking factor/fidelity for a
universal $N\rightarrow M$ cloner and generalised its operation to a more
general case of mixed and/or entangled input states which belong to the
symmetric subspace. Furthermore we have established the connection between
optimal quantum state estimation and optimal quantum cloning which allowed
us to extend the validity of the optimal state estimation methods~\cite{mapo}
to inputs of the above form.

We would like to express our special thanks to J.~I.~Cirac, who helped us to 
clarify the ideas presented in the paper.

It is a pleasure to thank C.H.~Bennett, V.~Bu\v{z}ek, 
 D.~DiVincenzo,
N.~Gisin, M.~Palma, S.~Popescu and R.~Werner for helpful discussions.
 We acknowledge support from the
Black Cow Caf\'{e}, which  provided an excellent working environment
for numerous discussions on quantum cloning.

This work was supported in part by the European TMR Research Network
ERP-4061PL95-1412, Hewlett-Packard, The Royal Society of London and
Elsag-Bailey, a Finmeccanica Company.

\end{document}